\documentclass[letterpaper]{sig-alternate}

\usepackage{ifpdf}
\ifpdf
\else
\fi

\usepackage{array}
\usepackage{url}
\usepackage{color}
\usepackage[colorlinks, citecolor=black, urlcolor=black, linkcolor=black]{hyperref}
\usepackage[small]{caption}

\begin{document}

\title{Real-time Bidding for Online Advertising: Measurement and Analysis}

\numberofauthors{1} \author{
\alignauthor
Shuai Yuan, Jun Wang, Xiaoxue Zhao\\
\affaddr{Department of Computer Science, University College London}\\
\email{\{s.yuan, j.wang, x.zhao\}@cs.ucl.ac.uk} }


\maketitle

\begin{abstract}

  The real-time bidding (RTB), aka programmatic buying, has recently become the fastest growing area in online advertising. Instead of bulking buying and inventory-centric buying, RTB mimics stock exchanges and utilises computer algorithms to automatically buy and sell ads in real-time; It uses per impression context and targets the ads to specific people based on data about them, and hence dramatically increases the effectiveness of display advertising. In this paper, we provide an empirical analysis and measurement of a production ad exchange. Using the data sampled from both demand and supply side, we aim to provide first-hand insights into the emerging new impression selling infrastructure and its bidding behaviours, and help identifying research and design issues in such systems.  From our study, we observed that periodic patterns occur in various statistics including impressions, clicks, bids, and conversion rates (both post-view and post-click), which suggest time-dependent models would be appropriate for capturing the repeated patterns in RTB. We also found that despite the claimed second price auction, the first price payment in fact is accounted for 55.4\% of total cost due to the arrangement of the soft floor price. As such, we argue that the setting of soft floor price in the current RTB systems puts advertisers in a less favourable position. Furthermore, our analysis on the conversation rates shows that the current bidding strategy is far less optimal, indicating the significant needs for optimisation algorithms incorporating the facts such as the temporal behaviours, the frequency and recency of the ad displays, which have not been well considered in the past.

\end{abstract}

\section{Introduction}

Online advertising is one of the most fast growing area in IT industry. Over the past 10 years, its revenue has increased from \$6.0 billion in 2002 to \$36.6 in 2012, with a compound annual growth rate of 19.7\%. The display-related advertising revenues totalled \$12 billion in 2012, with a 9\% increase from 2011 \cite{iab}. In display advertising the most significant concept in recent years is real-time bidding (RTB), or programmatic buying, where advertisers have the ability of making decisions for every impression (auction). In this paper we report interesting findings from field studies in a production ad exchange, and discuss related research challenges. RTB is a demand-side oriented concept. However, we use data from both advertisers and publishers for a better understanding of the bidding behaviour.

\begin{figure}[t]
	\centering
	\includegraphics[width=.5 \textwidth]{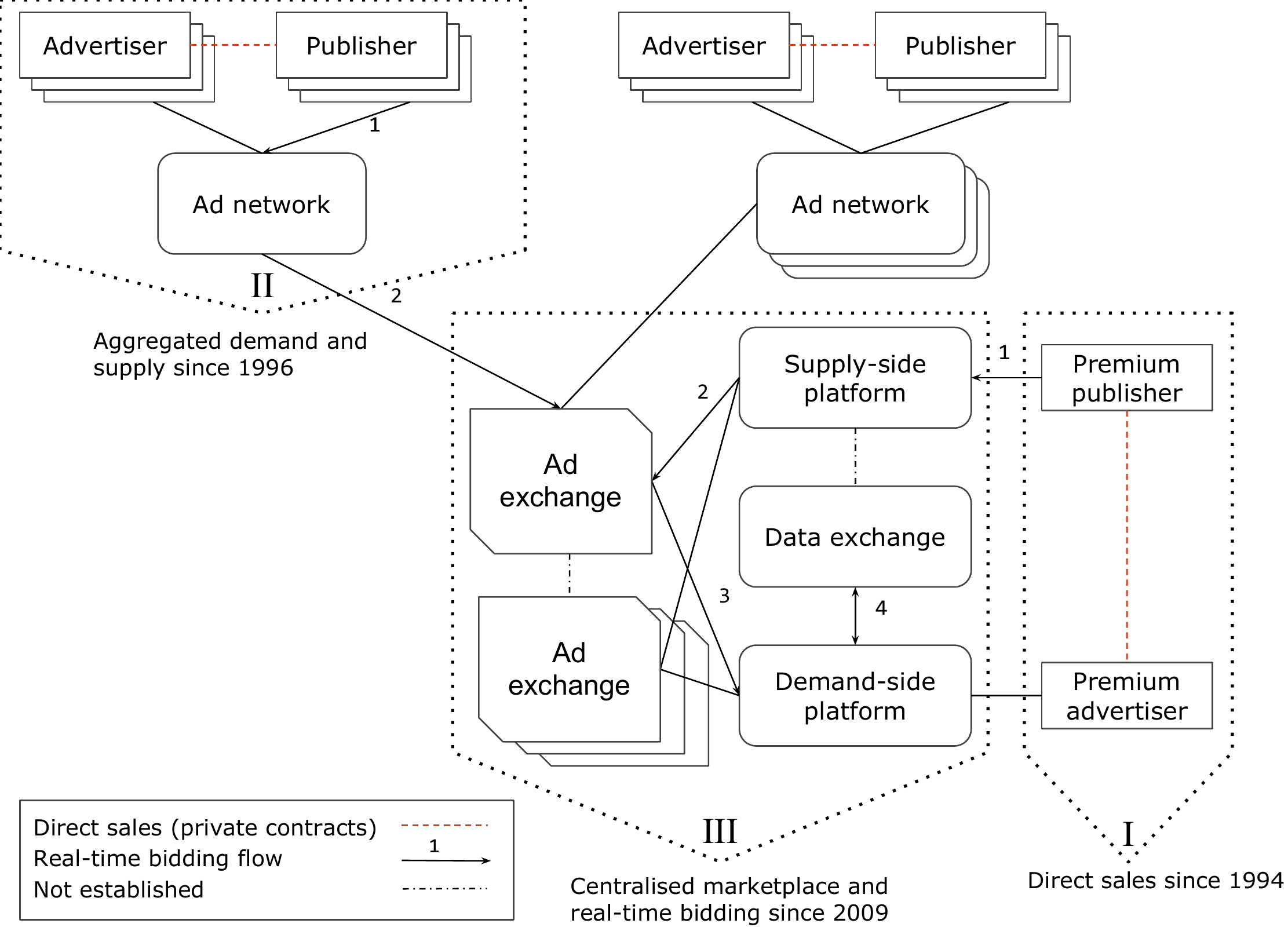}
	\caption{A brief illustration of the history and structure of display online advertising. Ad networks were created to aggregate advertisers and publishers. Ad exchanges were created to resolve the unbalance of demand and supply in ad networks. Premium advertisers and publishers now choose to work with ad exchanges through demand-side platform (DSP) and supply-side platform (SSP) to take the advantage of real-time bidding (RTB). The arrows with number describe the process of RTB: an impression is created then passed to ad exchanges; advertisers are contacted through DSPs; advertisers choose to buy 3rd party data optionally. Then following reversed path, bids are return to the ad exchange, then the SSP or ad network, and winner's ad will be display to the user on the publisher's website. The link between SSP and data exchange, as well as those among ad exchanges, are potentially useful however not widely adopted in current marketplace.}
	\label{fig-history_structure}
\end{figure}

We illustrate briefly the structure and history of display online advertising in Figure~\ref{fig-history_structure}. Throughout the paper we used advertiser/bidder/buyer/demand side interchangeably to comply with the industrial tradition. Since an advertiser rarely complete against himself, we consider a campaign is equivalent to an advertiser in our study. The case is similar for publisher where we used seller/supply side/website interchangeably. 

Before the emergence of RTB in 2009 (i.e., announcement of support by major ad exchanges \cite{google2011arrival}), the display advertising market was mainly divided by premium contracts (since 1994), which take more than 40\% of impressions, and ad networks (since 1996), which take the rest of impressions that were usually referred to as remnant. With premium contracts, publishers negotiate and make deals with advertisers directly, c.f. Block I in Figure~\ref{fig-history_structure}. Advertisers usually propose to buy certain amount of impressions from the given placements, regardless of the identities of users, when and how many times they have seen the ad, and so on. The purchase is totally about impressions and relies on the reputation of the publisher and reported audience profiles \cite{de2010predicting, bilenko2011predictive, abramson2012toward}. At the other end, the publisher needs to guarantee the delivery of impressions that have been agreed upon otherwise a penalty fee would be incurred \cite{roels2009dynamic, vee2010optimal}. The pricing model used in contracts are mostly cost-per-mille (CPM). Since advertisers have no control over the inventories or users, it is more difficult to deploy goal-driven campaigns (e.g booking a ticket) than branding ones (e.g announcing a new product). These contracts are sometimes called guaranteed delivery display advertising \cite{bharadwaj2010pricing}.

With ad networks, publishers register placements and offer impressions from them for sale. Impressions are largely sold using the second price auction in ad networks. Advertisers (or their delegates) also register with ad networks to participate in auctions. C.f. Block II in Figure~\ref{fig-history_structure}. However the impressions in ad networks are non-guaranteed, as opposed to premium contracts.

It is the ad network's responsibility to understand the webpage and the user, and to select advertisers based on their pre-defined targeting rules. The understanding of webpage is usually referred to as contextual advertising, where ad networks crawl, parse, and extract keywords which summarise the target. Advertisers bid on these keywords which is very similar to sponsored search \cite{ribeiro2005impedance, yih2006finding, anagnostopoulos2007just, wu2008keyword}. A more advanced approach is to learn a model including various features of webpages, which could then be used to compute a relevance score of advertisers' targeting criteria \cite{lacerda2006learning, broder2007semantic, chakrabarti2008contextual}. The understanding of user is usually referred to as behaviour targeting, where ad networks utilise the browsing history of a specific user to infer his interests, as well as geographical location, local time, etc for target matching \cite{yan2009much, provost2009audience, wu2009probabilistic}.

In ad networks advertisers largely adopt the cost-per-click (CPC) or cost-per-acquisition (CPA) pricing models where they only pay when certain goal is achieved. These choices reduce their risks thus are good for goal-driven campaigns. But then it is ad networks' responsibility to optimise to maximise clicks or conversions. In order to take the measurement of performance into account, ad networks usually employ the generalised second price auction (GSP) \cite{edelman2005internet} which allow them to apply bid biases (e.g the quality score) that usually weight the historical clickthrough rate (CTR) or conversion rate (CVR) heavily.

For the publisher side, an important research topic is to allocate impressions to premium contracts and ad networks. If chooses contracts, the publisher also needs to decide which specific contract to fulfil when multiple contracts having overlapping targeting rules. It is possible that ad networks bring in more revenue, but if a publisher sends too many impressions to this revenue channel and fails to fulfil a contract, he needs to pay the good-will penalty. In \cite{vee2010optimal} the authors used a two-phase models to sample, compute a compact allocation plan, and assign impressions online to advertisers who submit contracts with overlapping targeting rules. This balancing problem was discussed in \cite{roels2009dynamic} by using a certainty equivalent control heuristic to show the necessity of using both channels to reach the global maximum revenue. In \cite{feldman2010online} it was generalised as an online stochastic optimisation problem where given a set of resources, demands for resource arrive online with associated properties. Given a general priori about the demands, one has to decide whether and how to satisfy a demand when it arrives. The goal is to find a valid assignment (strategy) with the maximum total payoff.

When there were more and more ad networks, a problem led to the birth of ad exchanges: the excessive impressions in some ad networks. It is preferable to have more demand than supply because intense competition leads to higher revenue of both ad networks and publishers. However, when there are plenty of impressions unsold, ad networks try hard to find buyers. Besides, a common practice for advertisers was to register with multiple ad networks to find cheap inventories, or at least to find enough impressions within their budget constraints. They only found that managing numerous channels difficult and inefficient (e.g how to split the budget). Ad exchanges, like Google AdX, Yahoo! Right Media, Microsoft ad exchange, were created to address this problem by connecting hundreds of ad networks together, c.f. Block III in Figure~\ref{fig-history_structure}. Advertisers now have a higher chance to locate enough impressions with preferred targeting rules; publishers may received higher profit, too, because of more bidders potentially.

There are new research problems introduced by ad exchanges. In the pioneer work of \cite{muthukrishnan2009ad} the author discussed several issues including the truthfulness of auctions \cite{muthukrishnan2008internet}, callout optimisation \cite{chakraborty2010selective}, arbitrage bidding and risk analysis, etc. The most significant feature introduced by ad exchanges is real-time bidding, which queries bidders for a bid for every impression. Along with the query, ad exchanges send meta data of the webpage and the user. The exposure of this meta data enables advertisers to switch from the inventory-centric to user-centric optimisation. There are also noteworthy attempts to introduce concepts from finance market \cite{wang2012selling}. These attempts will make the ad exchange more mature and attractive.

When advertisers want to take advantage of RTB, they work with ad exchanges through 3rd party platform that are usually referred to as demand-side platform (DSP). DSPs are delegates of advertisers that answer bidding requests and optimise campaigns at the impression level. Comparing with ad networks, the advantages of using DSPs are: 1) advertisers do not need to manage their registration with many ad networks; 2) they can optimise at a finer granularity and a higher frequency because of local impression logs instead of aggregated reports from ad networks; 3) DSPs are also more customisable to better suit advertisers' goals. For example, advertisers traditionally set a frequency cap on their campaigns (the maximum times to display the ad to the same user). Now the cap can be applied to user groups, or even a single user for optimal efficiency.

At the other end, supply-side platforms (SSP) were created to serve publishers. Similarly, SSPs provide a central management console with various tools for publishers' ultimate goal: the yield optimisation. For example, SSPs allow publishers to set a reserve price for a specific placement, or even against a specific advertiser. Some SSPs also allow publishers to have preference over bidders (bid bias).

In Figure~\ref{fig-history_structure} arrows with number describe the process of RTB:
\begin{enumerate}
	\item An impression is created on publisher's website;
	\item The bidding request is sent to ad exchanges through ad network or SSP;
	\item The ad exchanges query DSPs for advertisers' bids;
	\item The DSP could contact data exchanges for 3rd party user data;
	\item The bid is generated and submitted; winner will be selected at ad exchanges, then at SSP. Following the reversed path the winner's ad will be displayed on publisher's website to the specific user.
\end{enumerate}

Note that ad exchanges normally only query DSPs for bids, not ad networks. Also note that there are two types of links missing from the current marketplace. We find these links interesting and potentially useful. Firstly, SSPs normally do not contact data exchanges for 3rd party user data. However, if publishers could understand and model specific users, they will be in a better position of revenue optimisation (e.g setting up an optimal reserve price for this auction based on the forecast of bidding activity with user data), although such query would incur cost. Secondly, ad exchanges normally do not send impressions to each other even if they remain unsold. This is largely due to business considerations and it leads to the fact that DSPs and SSPs register with multiple ad exchanges. A unified and interconnected marketplace is what we find attractive and it is certainly good for the whole eco-system.

In the remaining of this paper we represent our empirical study on a production ad exchange. We analyse the various datasets from both demand and supply side and report research challenges introduced by ad exchanges and RTB.

\begin{figure}[t]
	\centering
	\includegraphics[width=.5 \textwidth]{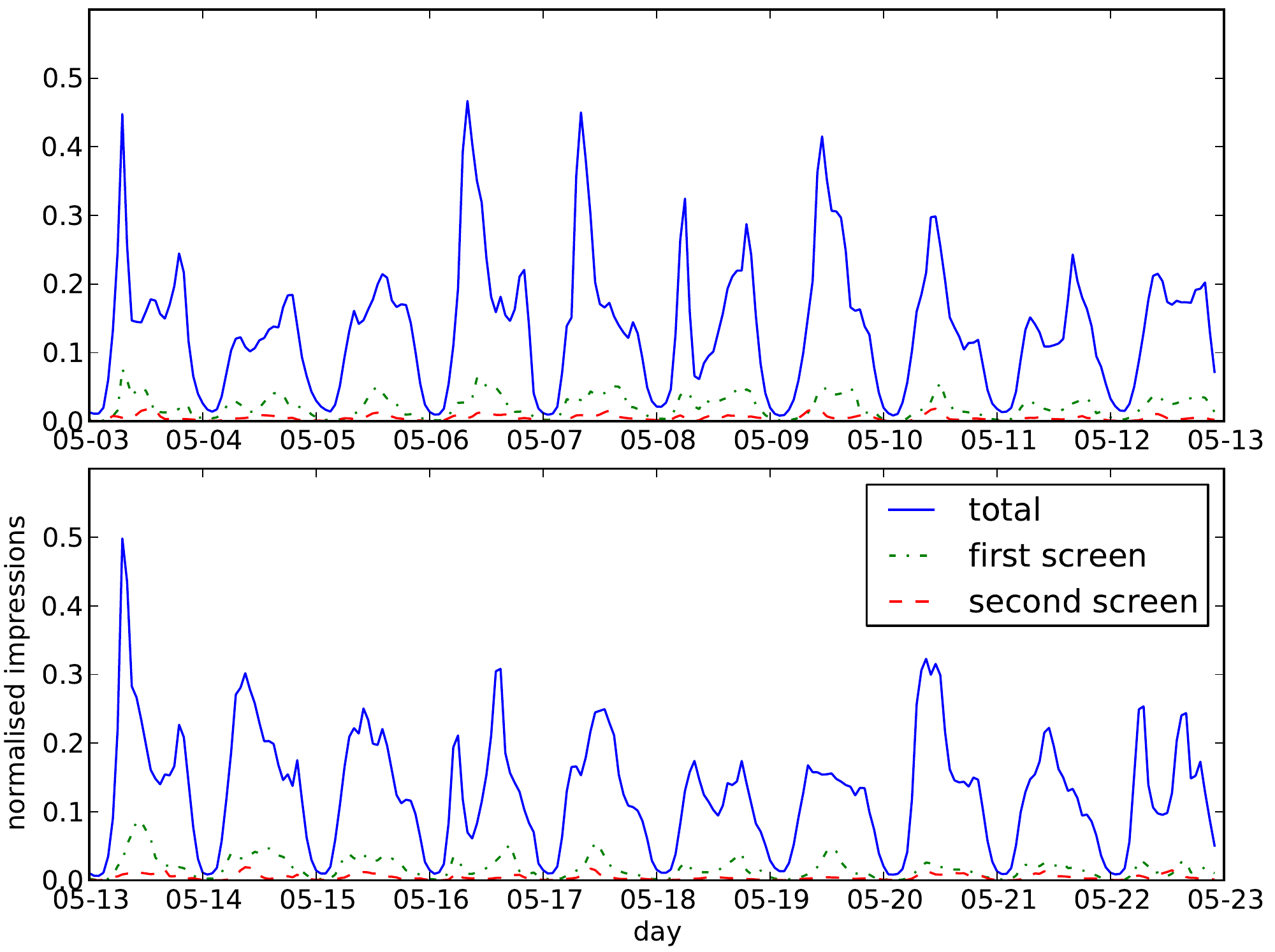}
	\caption{The normalised impression time series snippet by hour. A clear daily pattern and less clear weekly pattern, that less visitors come to the website, can be observed. The data is from a single website but the pattern is very similar for all websites we evaluated. The green dotted line denotes the impressions from the front page; the red segmented line denotes the impressions from the second-level pages (i.e. those are linked to directly from the front page).}
	\label{fig-imps_time_series}
\end{figure}

\begin{figure}[t]
	\centering
	\includegraphics[width=.5 \textwidth]{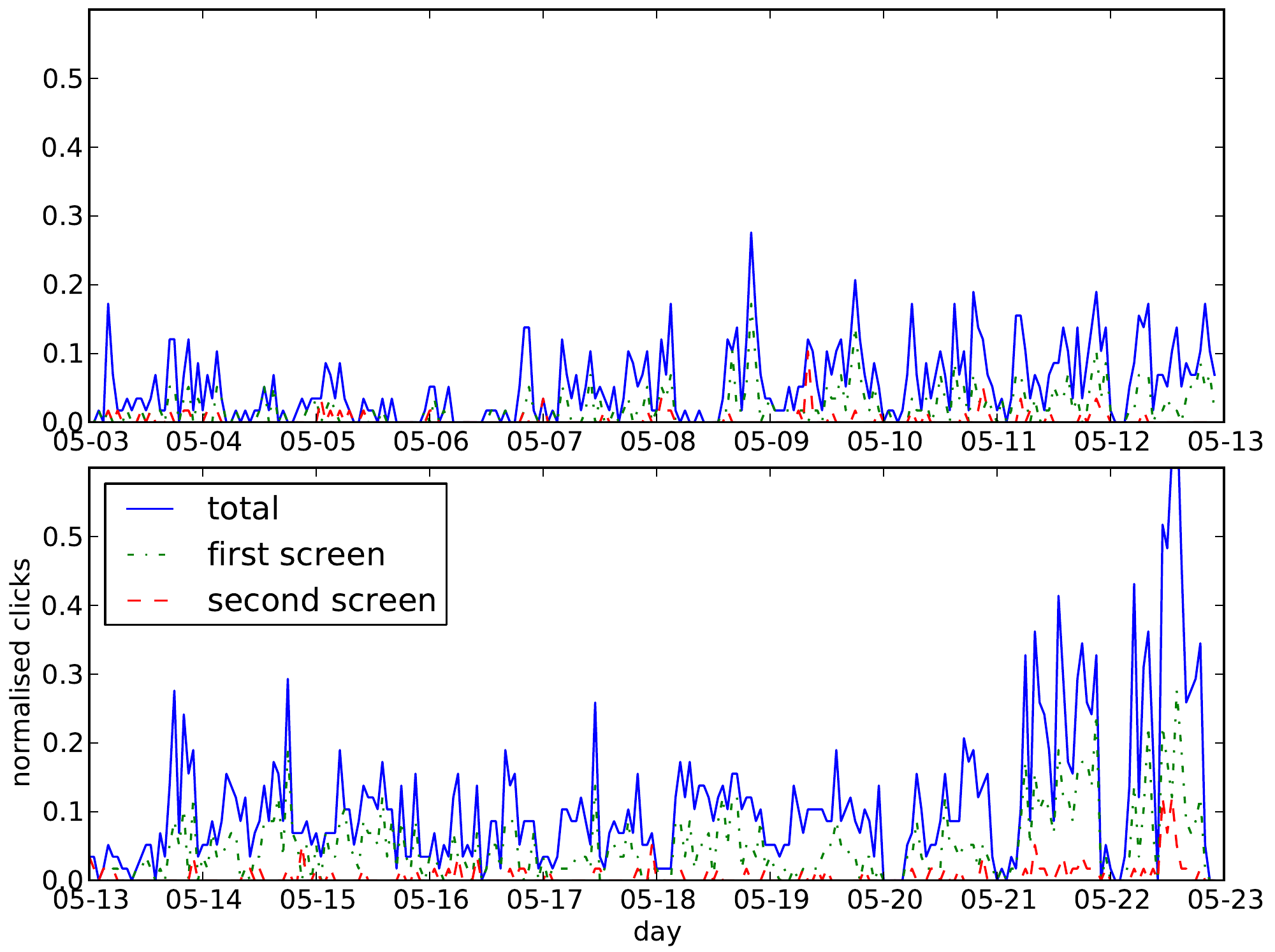}
	\caption{The normalised click time series snippet by hour from the same website as of Figure~\ref{fig-imps_time_series}. A clear daily pattern with lots of noises exists, too. Other websites we investigated had the very similar pattern.}
	\label{fig-clicks_time_series}
\end{figure}

\section{The empirical study of RTB}

We first report the statistics of dataset used in the empirical study and represent some general analysis on bidders and bids. Then we discuss two specific aspects (bidding behaviour, frequency and recency factor) and associated problems (floor price detection, daily pacing, and selective bidding). 

\subsection{Dataset and the System's Properties}

We conducted our study in a production ad exchange based in the UK. Datasets were created both for demand and supply side. For advertisers, we sampled impression, click, and conversion logs from Feb to May 2013. In total we sampled 52,850,635 impressions, 72,958 clicks, and 37,978 conversions. Note that we sampled the conversions only and back traced associated clicks and impressions to make the dataset self-contained (i.e. any impressions and clicks associating with a sampled conversion would be in the dataset). Since these datasets were sampled, we also used various reports (e.g analytical performance, site domain, attributed conversions, frequency/recency distribution, and so on) to understand the overall performance of advertisers.

We also recorded auction logs for placements from registered publishers of the ad exchange. In total we sampled 12,965,119 auctions from 50 placements, ranging from 14 Dec 2012 to 24 May 2013. These placements are from 16 websites of different categories, including news, finance, pc \& console games, gadgets, sports, entertainment, and so on. Since auction logs were sampled, we also used various reports to understand the overall performance of publishers.

In the following of this section we report some statistics that we found interesting in the study.

\subsubsection{Periodic Patterns}

\begin{figure}[t]
	\centering
	\includegraphics[width=.5 \textwidth]{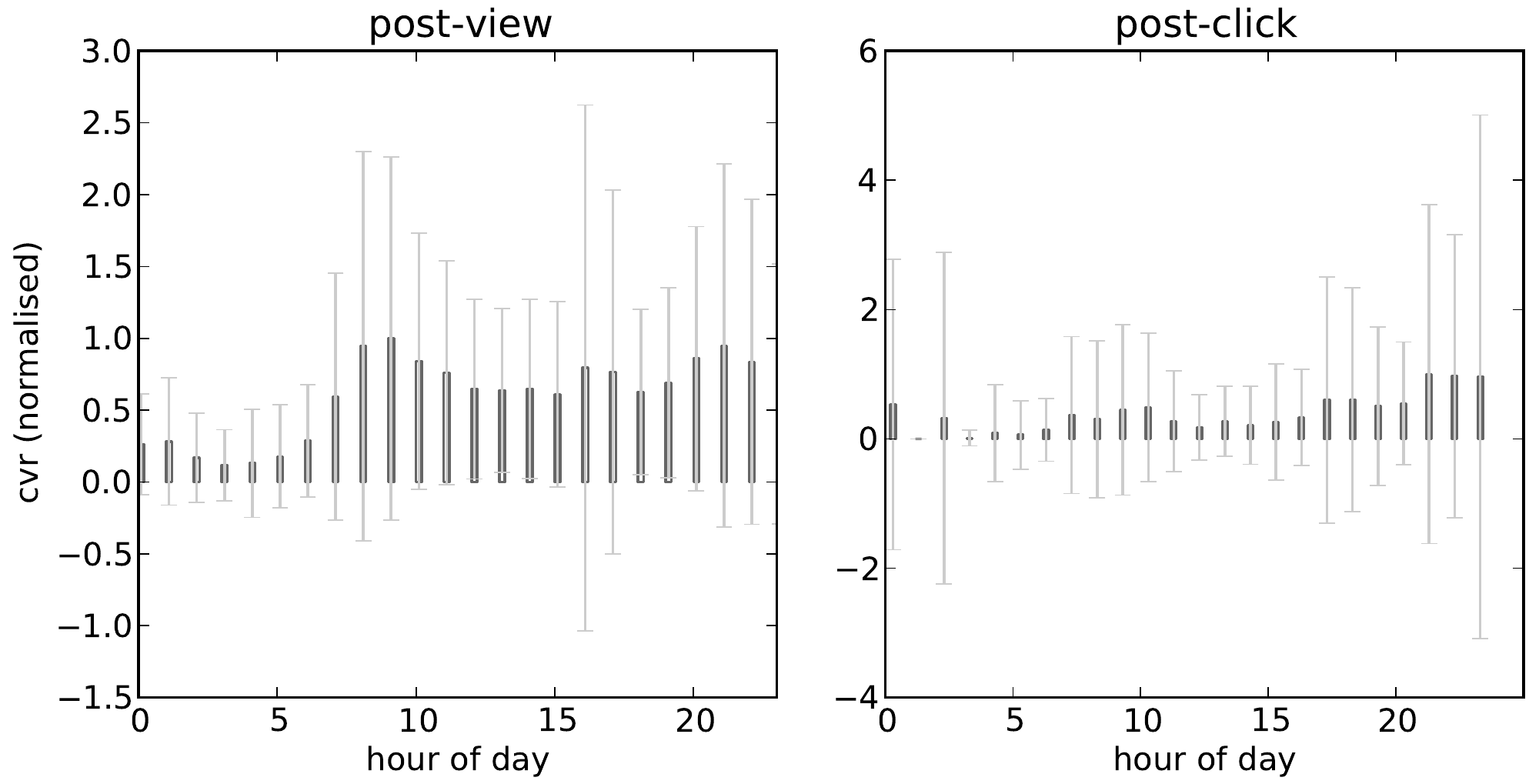}
	\caption{The distribution of normalised post-view and post-click CVR against time. The plot was created using all conversions in the demand-side dataset. The error bar shows the standard deviation. The plot shows that the CVRs were higher during daytime.}
	\label{fig-cvr_hour_hist}
\end{figure}

\begin{figure}[t]
	\centering
	\includegraphics[width=.5 \textwidth]{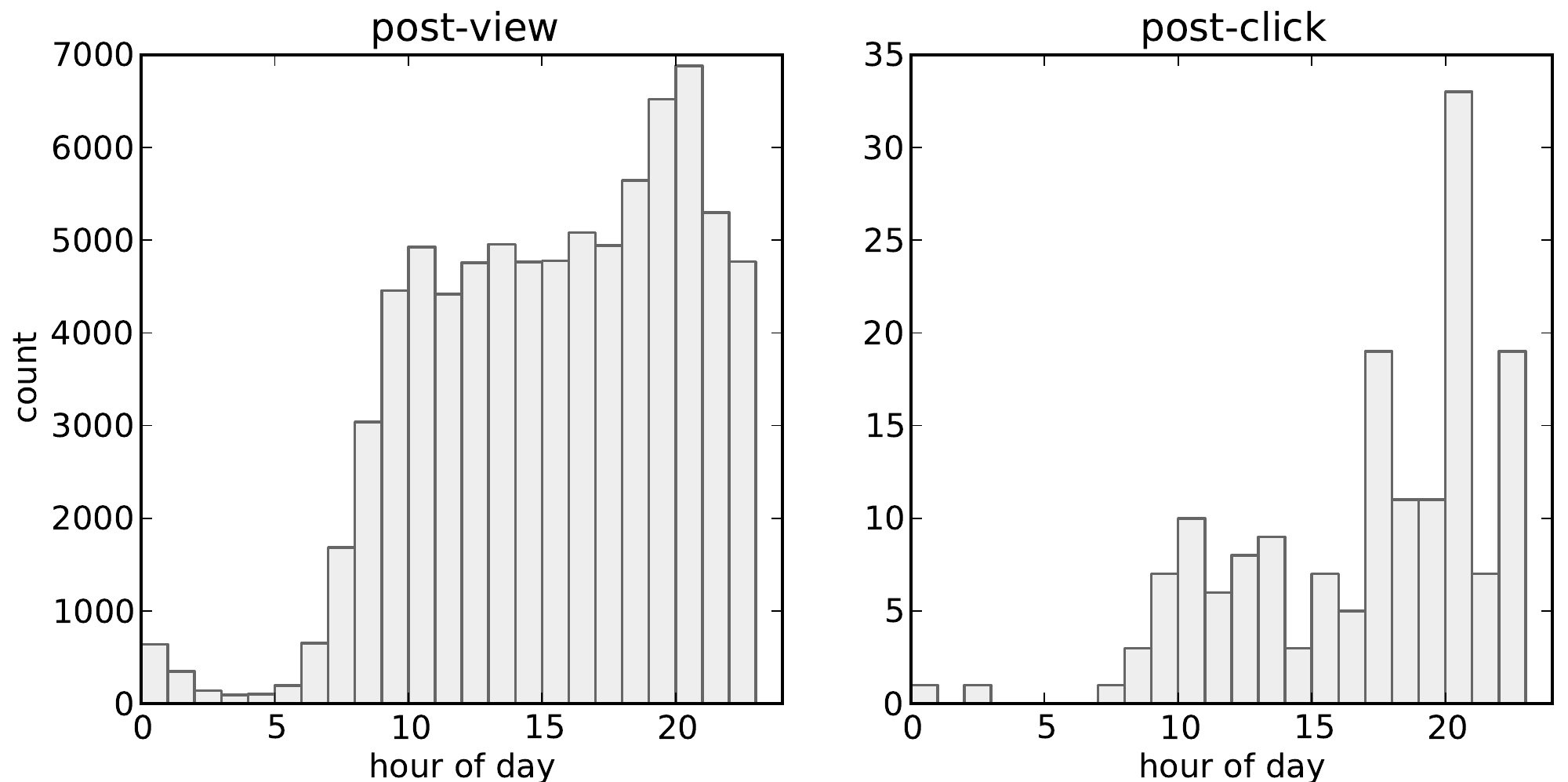}
	\caption{The distribution of number of post-view and post-click conversions against time. The plot was created using all conversions in the demand-side dataset.}
	\label{fig-conv_hour_hist}
\end{figure}

Due to the common daily routine of human there are always periodic patterns of our activity. In Figure~\ref{fig-imps_time_series} and Figure~\ref{fig-clicks_time_series} we can observe clear daily patterns of both impressions and clicks from a single website. For number of impressions, there was also a weak weekly pattern that during weekends less visitors come to the website. There was also a daily pattern for conversions: during daytime the numbers of conversions were larger and the CVRs were higher compared to sleeping hours, c.f. Figure~\ref{fig-conv_hour_hist} and Figure~\ref{fig-cvr_hour_hist}.

Note there are two types of conversions: \textit{post-click} that the user saw an impression, clicked, and finally converted (while remaining on the advertiser's website and without any interruption); \textit{post-view} that the user saw an impression, but did not click, but still converts (by visiting the website later on directly). These two types were plotted separately.


\subsubsection{The Impact of the Soft Floor Price}

\begin{figure}[t]
	\centering
	\includegraphics[width=.45 \textwidth]{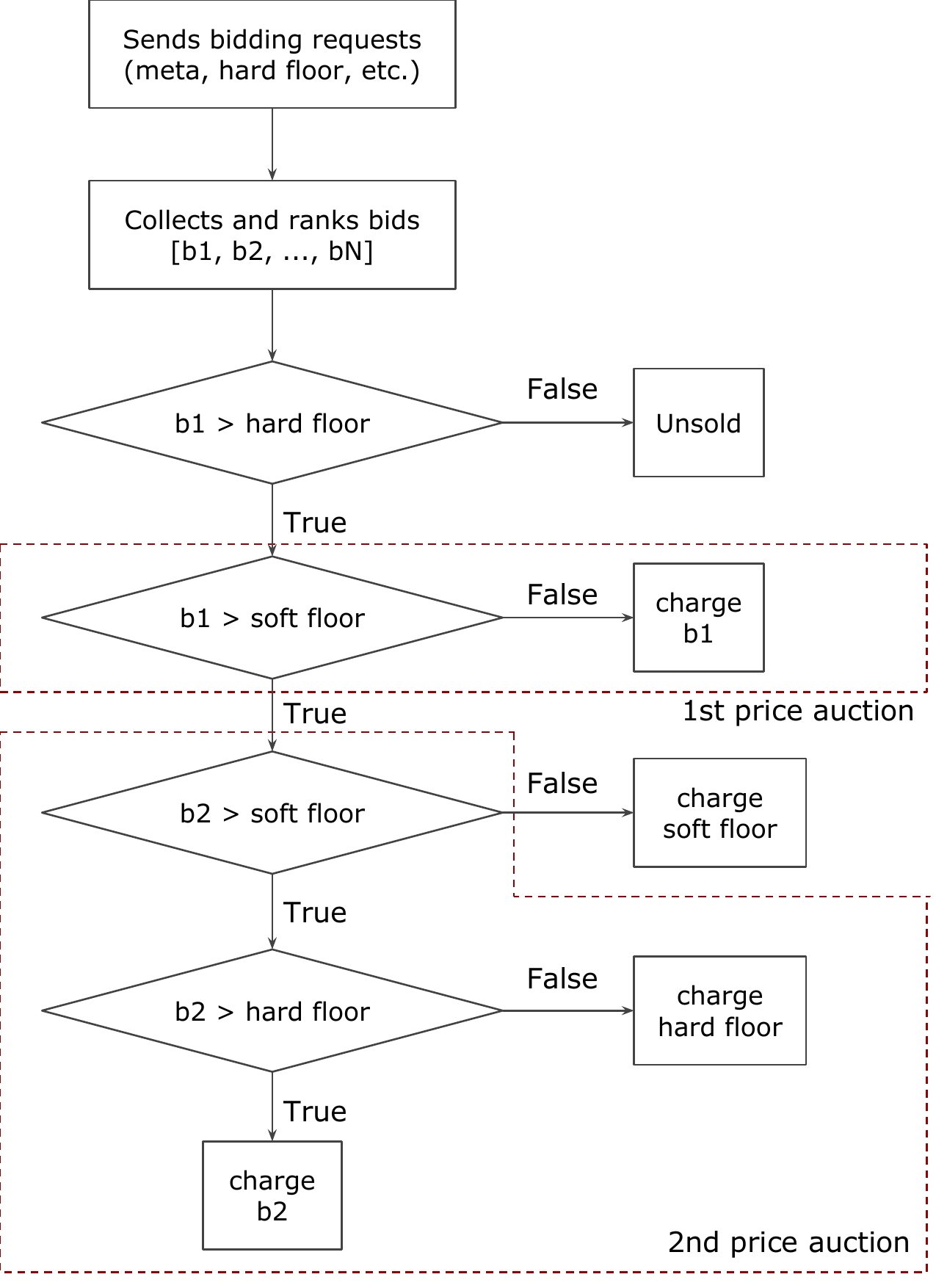}
	\caption{An illustration of the auction process in modern ad exchanges. The auction mechanisms are mixed and extended by introducing floor prices. The soft floor price is also in popularity but the impact of such mixture is largely unaware of. The complicated setting puts advertisers in an unfavourable position and could damage the eco-system.}
	\label{fig-auction_process_with_floor_prices}
\end{figure}

In modern ad exchanges publishers are usually given the option to set a soft floor price $p_s$ along with the hard floor price (the traditional reserve price) $p_h$. The auction process is illustrated in Figure~\ref{fig-auction_process_with_floor_prices}. By setting a high soft floor price (e.g \$1000 CPM), the publisher can change the auction from 2nd price to 1st price auction, which always charges the winner the amount that he bids. Unlike the hard floor, advertisers sometimes are not aware the existence of a soft floor, especially in RTB environment.

We checked the ratio of the (effective) 1st price auction and the 2nd price auction in the dataset, using impression logs that include bid price and price paid. There was no explicit indicator of the auction mechanism being employed, thus we used the simple heuristics: when (bid price = price paid), we considered it 1st price auction. In total, there were about 40\% impressions that were bought in 1st price auctions. However, these impressions accounted for 55.4\% of total cost. The existence of soft floor prices and the heuristics used could be examined in Figure~\ref{fig-daily_pacing}. To our best extent of knowledge this mixture of auction mechanisms is largely unaware of. The complicated setting, as illustrated in Figure~\ref{fig-auction_process_with_floor_prices}, puts advertisers in a not favourable position and could damage the advertising eco-system. From the advertisers' perspective, it is worth exploring the floor prices setting by different publishers and act accordingly. For example, a winner may choose to lower his bid (while still winning) to reduce the cost in a high soft floor scenario (the first price auction). Meanwhile, the advertiser who could react to its competitors' moves fastest has a substantial advantage \cite{edelman2005internet}; but if its the second price auction, the exploration activities only incur unnecessary cost.

\subsection{Bidding Behaviours}

We now study the advertisers bidding behaviours influenced by the design of the RTB systems. 

\subsubsection{Bids' Distribution} 

A widely adopted assumption in optimal auction design research is that the private values of advertisers follow log-normal distribution \cite{myerson1981optimal, edelman2006optimal, xiao2009optimal, ostrovsky2009reserve}. Bids are then generated from the private values. However, in the dataset we found that the 1st highest bid did not show strong log-normal distribution properties, nor did the 1st and 2nd highest bids, or all bids.

\begin{figure}
	\centering
	\includegraphics[width=.5 \textwidth]{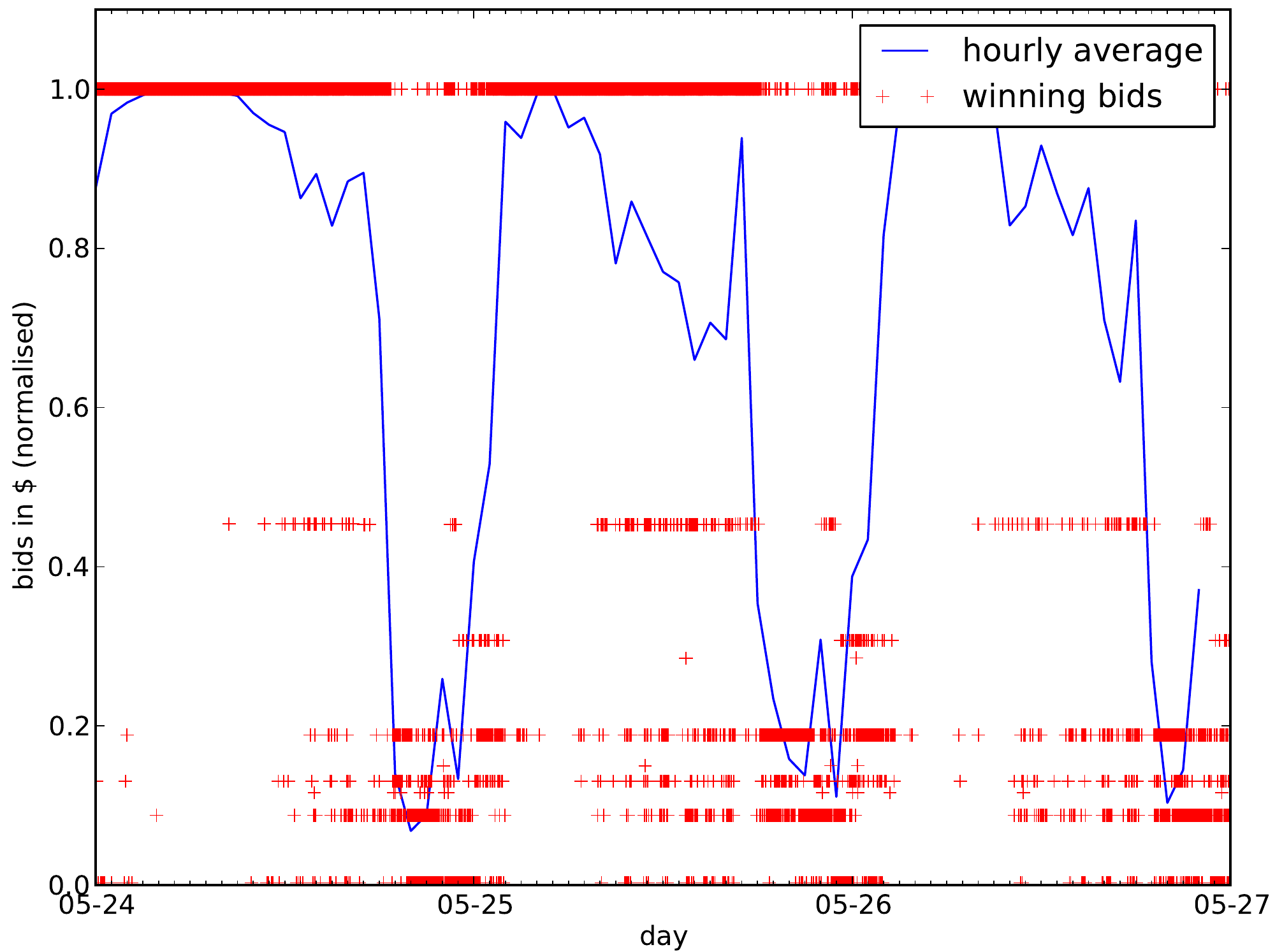}
	\vspace{-20px}
	\caption{The time series snippet of winning bids and its hourly average of a single placement. The hourly average series peaks around 6-8am every day when there are less impressions but more bidders.}
	\label{fig-winning_bids_time_series}
\end{figure}

We split auctions by hour-of-day since bids vary greatly throughout a day, c.f. Figure~\ref{fig-winning_bids_time_series}. From 50 placements and 160 days we obtained 192k <placement, hour> tuples. For each tuple we fit highest bids, highest + second highest bids, and all bids of every pair to Log-normal (actually to Normal after taking the logarithm) however the result is poor. There were less than 1\% accepted by either the Shapiro-Wilk test \cite{shapiro1965analysis} or the Anderson-Darling test \cite{anderson1952asymptotic}, regardless of the number of bidders or impressions in that hour.

\subsubsection{The Daily Pacing}

\begin{figure}[t]
	\centering
	\includegraphics[width=.5 \textwidth]{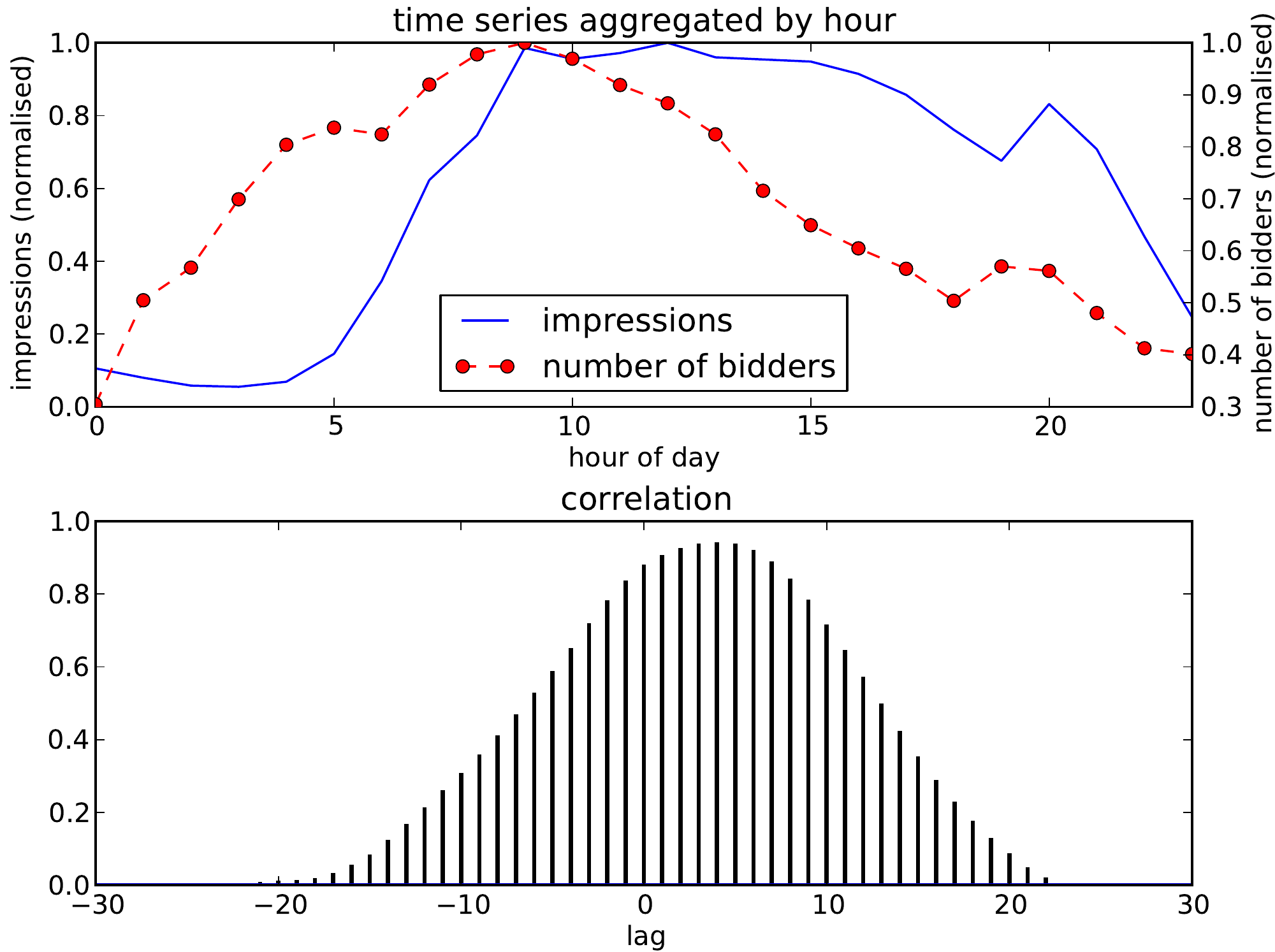}
	\caption{The distribution of number of bidders and impressions against hour-of-day. Their correlation plot shows a clear lag of when they reach the maximum in a day. This lag indicates the unbalance of supply and demand of the market in certain hours. Besides, the fact that there are more bidders in the morning may be due to the mixture of hour-of-day targeting and no daily pacing setup. The plots used 3 months worth of data sampled from a single placement. Note for some placements the lag was not very clear.}
	\label{fig-imps_bidders_corr}
\end{figure}

\begin{figure}[t!]
	\centering
	\includegraphics[width=.5 \textwidth]{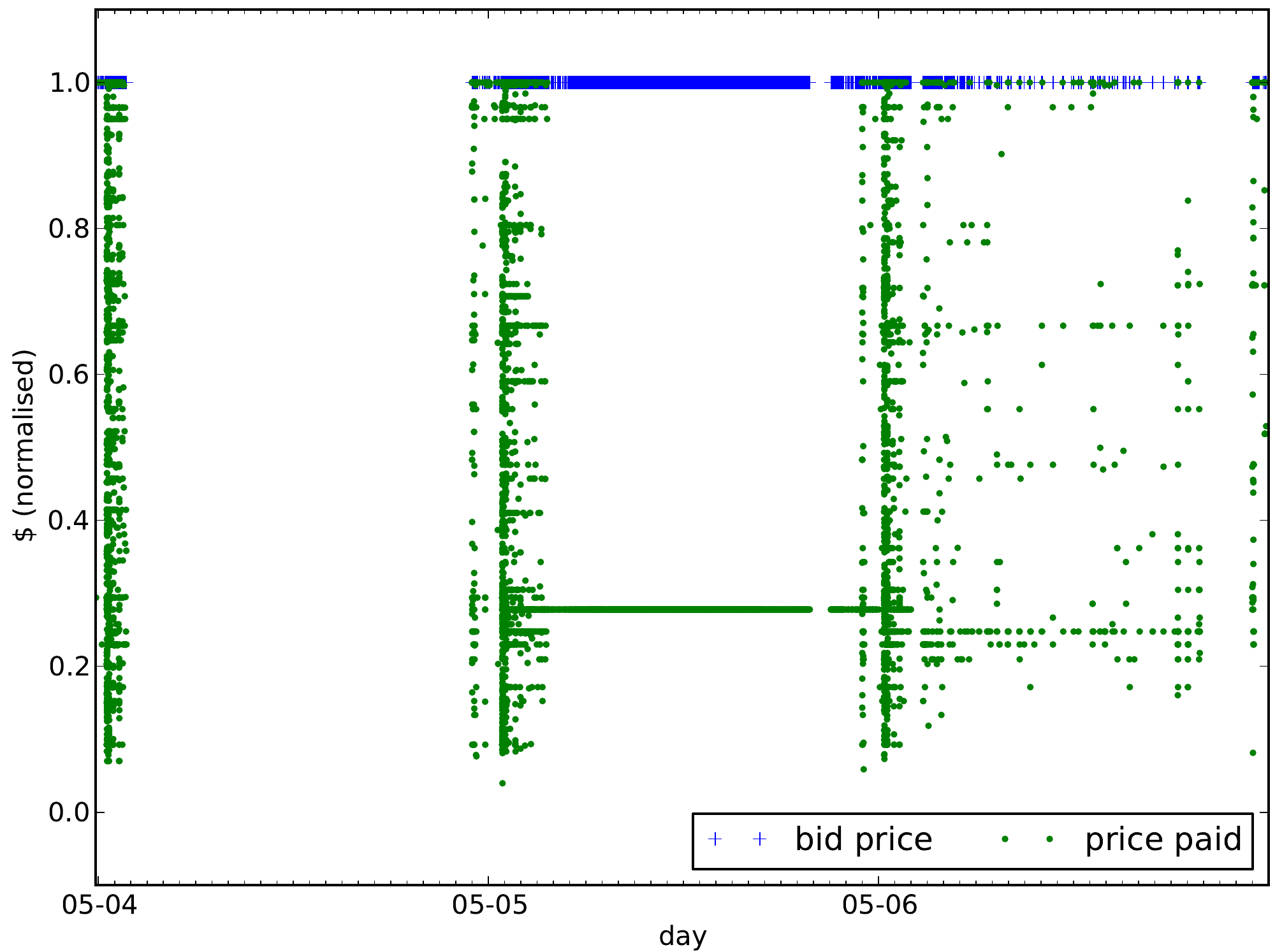}
	\caption{An interesting instance we found in the dataset: an advertiser switched from no pacing to even daily pacing. He was bidding at a flat CPM. With ad exchanges, the large amount of available inventories could deplete a daily budget quickly. However, even daily pacing is far from optimal because it does not consider the performance (e.g ROI) in different time slots. Note that in practice the pacing engine would spend slightly more in the beginning of the day to learn the available impressions and to calculate the spending speed, especially for campaigns with small budgets.}
	\label{fig-daily_pacing}
\end{figure}

The daily pacing refers to the way that advertisers spend their budget in a single day. Usually an advertiser submits a daily budget for his campaign, and choose from spending it evenly throughout a day (uniform pacing) and as fast as possible (no pacing). The no pacing setup may lead to premature stop easily, which means the budget depletes too quickly so advertisers cannot capture traffic later in the day, that may have high quality impressions. An instance of premature stop is illustrated in Figure~\ref{fig-daily_pacing} (day 1). The uniform pacing also suffers from the traffic problem: if high quality impressions appear in early part of the day, the pacing setup would not be able to capture all of them; if there is not enough traffic in the late part of the day, the pacing setup would not be able to spend all the budget (usually called under-delivery). In sum they are not good daily pacing strategies although being widely used in DSPs.

In Figure~\ref{fig-imps_bidders_corr} we plot the hourly mean of number of bidders and number of impressions, that were normalised respectively. There is a clear lag of when these two series reach their maximum in a day. Generally speaking, the number of bidders peaks in the morning but the number of impressions peaks afterwards. This lag indicates the unbalance of supply and demand of the market in certain hours. For example, there are more bidders competing over limited impressions in the morning, resulting in high winning bids as shown in Figure~\ref{fig-winning_bids_time_series}, which in turn costs more of advertisers. However, this higher cost does not necessarily lead to higher performance. From Figure~\ref{fig-cvr_hour_hist} we can see that both post-view and post-clicks CVR peak in the evening, which argues that intense bidding activities in the early hours are not reasonable.

This distribution of number of bidders throughout a day may be due to the mixture of hour-of-day targeting and no daily pacing setup. Most of advertisers wish to skip the last night hours because of the low CVR. Some of them may use the no daily pacing setup to avoid the risk of under-delivery as we discussed before. When these bidders start to bid in the morning, they win lots of impressions with high bids, and they quit as their budgets depletes quickly. We leave the validation of this hypothesis to the future works.

A reasonable alternative is the dynamic pacing against the performance. It is intuitively correct to spend more budget in hours that generate more clicks or conversions, and less in low performing hours.
%
%

Note that this problem is not the same as a typical budgeted multi-arm bandit problem that has been discussed extensively in the literature \cite{madani2004budgeted, deng2007bandit, tran2012knapsack}: 1) In this allocation problem the advertiser can only explore the current time unit; 2) There is potential over or under-delivery in a time unit due to the latency in the practical implementation. Therefore revising the remaining budget is required after every time step.




%
%


\subsection{Conversion Rates and Selective Bidding}

Considering the maintenance cost (e.g servers and bandwidth) it is not wise for an advertiser to submit a bid every time he receives a bidding request. Among various factors that he uses to decide to bid or not, the frequency and recency factor are important ones that have not been given enough attention to.

\subsubsection{The Frequency Factor}

The frequency factor (or frequency cap, FC) defines how many times ads (a.k.a. creatives) would be displayed to a single user. It can be applied to campaign groups, campaigns, and creatives separately. For example, given a campaign with 3 ads, an advertiser could set FC(campaign) = 6, FC(ad$_1$) = 2, FC(ad$_2$) = 3, and FC(ad$_3$) = 4 and these settings would work together. The same user would at most see ad$_1$ twice, ad$_2$ three times, ad$_3$ four times, and all ads displayed to him together would not exceed six times.

\begin{figure}[t]
	\centering
	\includegraphics[width=.45 \textwidth]{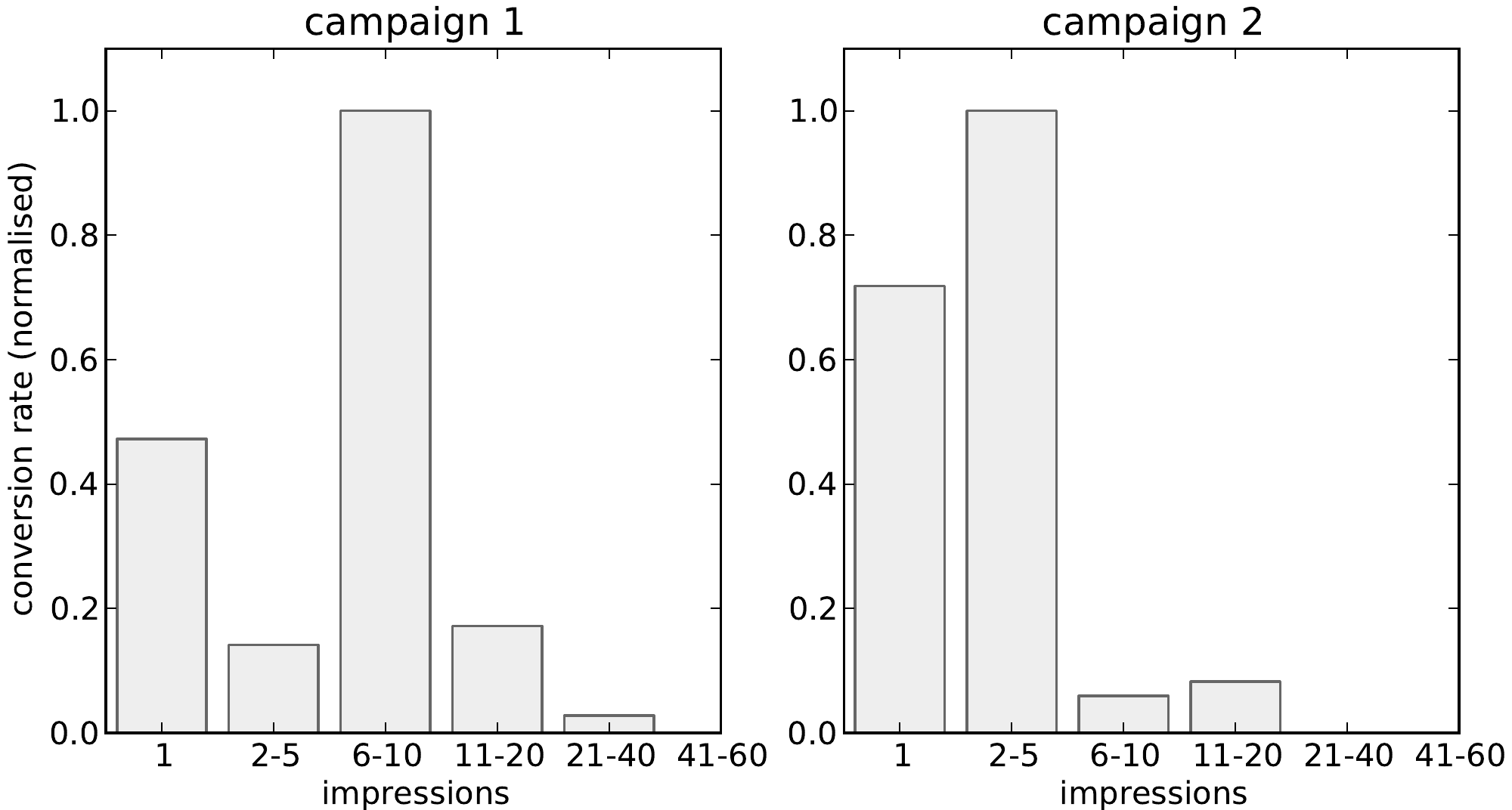}
	\caption{The frequency against CVR plot from two different campaigns. The campaign 1 from the left plot received the highest CVR with 6-10 impressions, which is also true for the cumulative CVR metrics. However the campaign 2 from the right plot received the highest CVR with 2-5 impressions. If the campaign 1 sets a frequency cap of 2-5 impressions, most of conversions would not be achieved. If the campaign 2 sets a frequency cap of 6-10, nearly half of impressions could be wasted.}
	\label{fig-frequency_cvr_bar}
\end{figure}

The frequency factor is normally set based on historical data and needs to be adjusted constantly during the flight time of the campaign, because different campaigns ask for very different FCs, as illustrated in Figure~\ref{fig-frequency_cvr_bar}. The campaign 1 from the left plot received the highest CVR with 6-10 impressions, which is also true for the cumulative CVR metrics. However the campaign 2 from the right plot received the highest CVR with 2-5 impressions. If the campaign 1 sets a frequency cap of 2-5 impressions, most of conversions would not be achieved. If the campaign 2 sets a frequency cap of 6-10, nearly half of impressions could be wasted. Therefore, a good FC setting is crucial to the efficiency of advertising.

Before setting an optimal FC, we need to find the right metrics to measure the efficiency of different FCs. See the example in Table~\ref{tbl-cumulative_cvr} where we compare popular metrics. Assume there are 100 users; the CPM is fixed at 10; the advertiser's conversion goal is worth 500. From the table we can see using different metrics could lead to very different decision. If the advertiser uses CVR as most of advertisers do, he would go for FC=3; if the advertiser cares more about the total number of conversions, he would use FC=5; if the advertiser cares more about CPA he would use FC=3, too; if the advertiser measures the performance against the profit and cost, he would go for FC=2.

Using FC = 2 seems the most profitable. However, choosing FC = 3 is reasonable, too, especially when considering the long term impact: FC = 3 gives more conversions at low cost, and once these users are attracted they have a higher chance of converting again in future.

\begin{table}[t]
	\centering
    \begin{tabular}{l|l|l|l|l|l}
	fc & cvr & (cumulative) cvr & convs & cpa  & roi  \\ \hline \hline
    1           & 0.0000 & 0.0000         & 0           & -    & 0.00 \\ \hline
    2           & 0.0150 & 0.0150         & 3           & 667  & \textbf{1.50} \\ \hline
    3           & 0.0067 & \textbf{0.0167}         & 2           & \textbf{600}  & 1.25 \\ \hline
    4           & 0.0025 & 0.0150         & 1           & 667  & 1.00 \\ \hline
    5           & 0.0020 & 0.0140         & 1           & 714  & 0.88 \\ \hline
    6           & 0.0000 & 0.0117         & 0           & 857  & 0.70 \\ \hline
    7           & 0.0000 & 0.0100         & 0           & 1000 & 0.58 \\ \hline
    \end{tabular}
    \caption {The comparison of different metrics against frequency caps (FC). Note \textit{cvr} and \textit{convs} are \textit{fc} specific, i.e. the extra value could be gained by increasing the frequency cap from the previous level to the current one. The \textit{cumulative cvr} is the CVR advertisers use: total conversions divided by total impressions. The \textit{cpa} gives the cost-per-acquisition. The \textit{roi} gives the return-on-investment based on the advertiser's conversion valuation. These values are calculated by assuming there are 100 users; the CPM is fixed at 10; the advertiser's conversion goal is worth 500. }
    \label{tbl-cumulative_cvr}
\end{table}

\subsubsection{The Recency Factor}

The recency factor (or recency cap, RC) helps to decide to bid or not based on how recently the ad was displayed to the same user. It also works at different levels including campaign groups, campaigns, and creatives. For example, an advertiser can set RC(campaign) = (1 hour) so that all ads from this campaign would be displayed to the same user only once in every hour. Similar to the frequency factor, RCs are useful to achieve the best advertising efficiency. For example, displaying ads intensively incurs high cost but little effect (or even getting people annoyed) for some campaigns (e.g financial services). Users need to think, compare, then make decisions. A better strategy is to display the same ad right after the \textit{thinking time} to remind users. However, for some campaigns (e.g booking flight) users would convert very quickly or not convert at all, which requires relatively intense advertising.

\begin{figure}[t]
	\centering
	\includegraphics[width=.45 \textwidth]{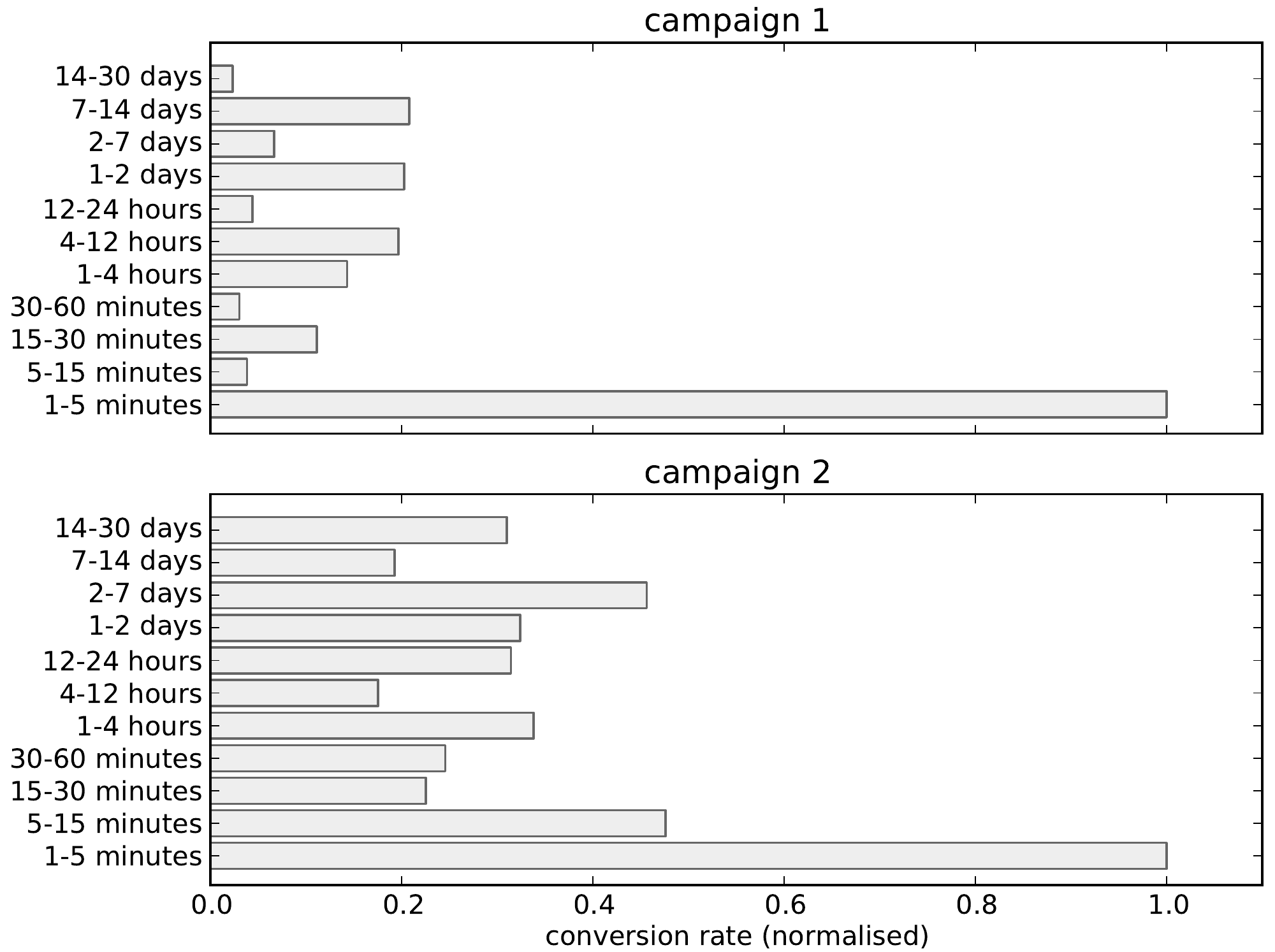}
	\caption{The recency factor against CVR plot from two different campaigns. Both campaigns show the highest CVR at the 1-5 minutes level. However for the campaign 2 the CVR is still not negligible after a long time (14-30 days). If the advertiser uses a more strict RC setting (e.g do not display ads to users who were first exposed 14 days ago) he could lose potential conversions. On the other hand, using a loose RC setting for the campaign 1 will only waste budget since the CVR is very low after 14 days.}
	\label{fig-recency_cvr_bar}
\end{figure}

Figure~\ref{fig-recency_cvr_bar} plots the RC against CVR of two different campaigns in the dataset. Both campaigns show the highest CVR at the 1-5 minutes level. However for the campaign 2 the CVR is still not negligible after a long time (14-30 days). If the advertiser uses a more strict RC setting (e.g do not display ads to users who were first exposed 14 days ago) he could lose potential conversions. On the other hand, using a loose RC setting for the campaign 1 will only waste budget since the CVR is very low after 14 days.

\begin{figure}[t]
	\centering
	\includegraphics[width=.45 \textwidth]{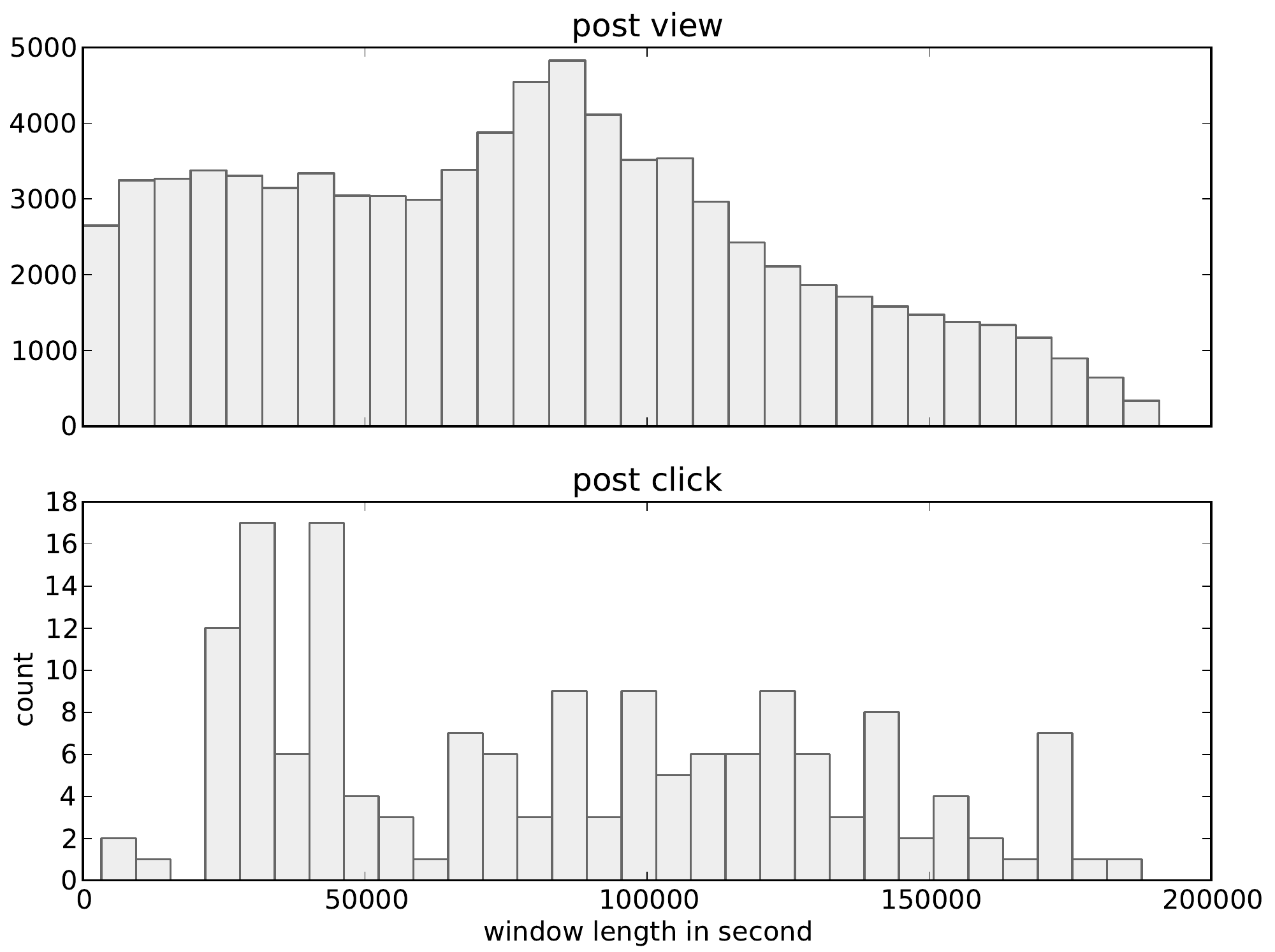}
	\caption{The histogram of conversion window lengths since the first exposure. Post-view and post-click conversions are drawn separately but are from the same campaign. The plot for the post-view conversions roughly distinguish people into two groups: \textit{impulse purchaser} who converted quickly and \textit{rational purchaser} who took time to consider after seeing the first ad. In order to maximise conversions, the advertiser would consider to set up RCs to target these two types of users. Interestingly the plot of post-click conversions suggests that the time needed to fill the form and complete the purchase varied a lot for different users (or they could leave the page open for a long while).}
	\label{fig-conv_window_length_hist}
\end{figure}

Another interesting observation is given in Figure~\ref{fig-conv_window_length_hist}. Similarly to the frequency factor, the analysis of efficiency of RCs requires metrics based on the understanding of advertising goal, which we do not repeat here.

The above analysis shows the importance of setting up proper FCs and RCs. At present these settings are at most at the creative level. With real-time bidding, advertisers can push them to a finer granularity: setting FCs and RCs for individual users. The individual FCs and RCs can then be used to decide to bid or not on a specific impression.

%
%

\section{Conclusion}

In this paper we introduced the history of real-time bidding and discussed research issues related to the demand side. In fact, RTB, ad exchanges, and DSP are conceptual ideas; what we really want to explore are the behaviour of advertisers and their delegates in the market, and the challenges brought by the impression-level bidding and user-centric bidding, distinguished from bulk buying and inventory-centric buying. Through analysis of datasets acquired from a production ad exchange, we discovered that floor price detection, daily pacing, and frequency/recency setting are problems not addressed. We explained their importance to advertisers however leave the development and evaluation of algorithms to the future works.

{
\bibliographystyle{acm}
\bibliography{bib}
}
\end{document}